\documentclass[aps]{revtex4}
\usepackage{amsmath}
\usepackage{amsfonts}
\usepackage{amssymb}
\usepackage{amstext}
\usepackage{graphicx}
\usepackage{epsfig}%
\usepackage{graphicx}%
\usepackage{color}
\def\be{\begin{equation}}
\def\ee{\end{equation}}

\def\bea{\begin{eqnarray}}
\def\eea{\end{eqnarray}}
\begin{document}
\title{Thermodynamics of the Pauli oscillators and Lee-Wick partners of the 
Standard model particles}
\author{Kaushik Bhattacharya}
\email{kaushikb@iitk.ac.ini}
\affiliation{Department of Physics, Indian Institute of Technology, 
Kanpur, Kanpur 208016, India}
\author{Suratna Das}
\email{suratna@prl.res.in}
\affiliation{Theoretical Physics Division, Physical Research Laboratory, 
Navrangpura, Ahmedabad 380009, India}
\begin{abstract}
The present article is about the statistical mechanics of non-trivial
field configurations. The non-trivial fields arise from the negative
sign of the commutators and the anticommutators of the bosonic and
fermionic field excitations. These kinds of fields were previously
studied by Pauli and Lee and Wick. The thermal distribution function
of the above mentioned fields are calculated in the article and using
the thermal distribution functions the energy density, pressure and
entropy density of the non-trivial field configurations are found
out. The results match exactly with a previous calculation done by
Fornal et. al. for higher derivative Lee-Wick theories showing a
deeper similarity with the earlier work. It is assumed that such kinds
of non-trivial fields may have existed in the early universe and may
have some cosmological relevance.
\end{abstract}
\maketitle
\section{Introduction}
For the last couple of decades there has been a series of papers
related to the Lee-Wick kind of field theories. The Lee-Wick
construction \cite{Lee:1969fy,Lee} originally was motivated to tackle
the problem of infinities in quantum field theories. By including
massive partners of the photon and other fermions present in the
theory of Quantum electrodynamics Lee and Wick could produce a theory
which was devoid of divergences. Later there has been a flurry of
activity concerning the connection of higher derivative theories and
the Lee-Wick construction \cite{Grinstein,Carone:2008bs}. All these
theories assumed the existence of some partners of the Standard model
particles. The partners of the Standard model fields arose out of
suitable redefinitions of fields in a higher derivative version of the
Standard model Lagrangian.  In these cases the propagators of the
partners of the Standard model fields behaved more like the
propagators one would get if they had a Lee-Wick like theory. There
has been at least one attempt \cite{Cai:2008qw} to use the concepts of
these Lee-Wick constructions in cosmology where the authors were able
to show the bouncing nature of the universe whose energy is dominated
by the energies of a scalar field and its Lee-Wick partner. In
Ref.~\cite{Fornal} the authors tried to formulate a possible
thermodynamic theory of particles which includes the Lee-Wick partners
using a method of statistical field theory previously formulated by
Dashen, Ma and Bernstein in Ref.~\cite{Dashen}.

In 1943, Pauli wrote a paper on an unusual field theory \cite{Pauli},
where the bosons appearing in the theory were accompanied by partners,
whose excitation modes were quantized with the wrong sign of the
commutators. The idea of proposing such an unusual theory came from an
urge to suppress ultraviolet divergences appearing in normal quantum
field theories. Most of the work of Lee and Wick \cite{Lee:1969fy,Lee}
utilize the concepts of Pauli, and his field excitations which will be
called Pauli oscillators in this paper. Pauli's paper did not have any
ingredients from a higher derivative theory.

There has not been any attempt to produce a statistical field theory
based on Pauli's work up till now. One of the important outcomes of
this article is to show that the results obtained on the
thermodynamics of a Lee-Wick ghost infested universe, as done in
\cite{Fornal}, matches exactly with the thermodynamic quantities of an
universe full of Pauli oscillators. We have included the fermionic
sector also, where the basic quantization prescription was given by
Lee and Wick \cite{Lee}. In this regard it should be noted that unlike
the fermions in Lee and Wick's theory the fermion quantization
presented in this article is simpler and the fermions do not have a
complex mass. These facts may have some bearing on the unitarity of
the theory but presently the theory predicts thermodynamic results
which are in agreement with the results obtained in
Ref.~\cite{Fornal}. 

The previous attempt to theorize the thermodynamic properties of a
universe which is constituted by the Standard model particles and
their possible Lee-Wick partners gave interesting results which may
have very important consequences in cosmology. In the present work we
use the basic commutation relations and number operator language as
used in the original paper of Lee and Wick, and Pauli \cite{Pauli},
and derive the expressions of the thermal distribution functions of
the Lee-Wick ghosts or the Pauli oscillators which accompany the
normal particles present in the Standard model. The distribution
functions of such highly non-trivial field configurations are
presented for the first time in this article.  The thermal
distribution function of the unusual field configurations are
different in nature from those of the standard Bose-Einstein or
Fermi-Dirac distributions. Using the new distribution functions we can
reproduce all the thermodynamic quantities previously calculated by
Fornal, Grinstein and Wise in Ref.~\cite{Fornal}.

The material in the present article is presented in the following
manner.  The next section discusses about the technique to find out
the thermal distribution function of the Lee-Wick partners. In section
\ref{eps} the energy density, pressure and entropy density of a gas
comprising of elementary particles and their unusual field partners
are calculated using the thermal distribution functions. The last
section \ref{conc} summarizes the important points of the article.
\section{Distribution functions of the Pauli oscillator excitations and 
Lee-Wick fields}
\label{distrb:s}
A free quantum field theory can be thought of to be made up of an
infinite number of linear harmonic oscillators in the momentum space
oscillating with frequencies $\epsilon({\bf p})=\sqrt{{\bf p}^2+m^2}$
where ${\bf p}$ is the 3-momentum of the oscillator excitations which
can take infinite values, and $m$ is the mass of the excitations over
the vacuum. For each of the oscillators, corresponding to any ${\bf
  p}$, there corresponds an annihilation operator $a({\bf p})$ and a
creation operator $\bar{a}({\bf p})$. The operator $\bar{a}({\bf p})$
is defined as $\bar{a}({\bf p})\equiv\eta^{-1} a^\dagger({\bf p})
\eta$\,, where $\eta$ is the metric on the Hilbert space of the
oscillators which is $+1$ for most of the cases, but need not be so in
general. The oscillators can be bosonic, if the only non-vanishing
commutators are of the form $\left[a({\bf p})\,,\,\bar{a}({\bf
    k})\right]=\delta^3({\bf p}-{\bf k})$\,, or fermionic if the only
non-vanishing anti-commutators are of the form $\left\{a({\bf
  p})\,,\,\bar{a}({\bf k})\right\}=\delta^3({\bf p}-{\bf k})$\,. In
the above cases the metric on the Hilbert spaces of the bosonic or
fermionic oscillators are both $+1$. But the above mentioned
commutation relations do not exhaust all the possibilities of
quantizing the oscillators. Instead of the usual bosonic commutation,
if we assume
\begin{eqnarray}
\left[a({\bf p})\,,\,\bar{a}({\bf k})\right]=-\delta^3({\bf p}-{\bf k})\,,
\label{bquant}
\end{eqnarray}
then also we get oscillator like spectrum where the eigenvalues of 
the number operator \cite{Lee}:
\begin{eqnarray}
N({\bf p})\equiv -\bar{a}({\bf p})a({\bf p})
\label{N}
\end{eqnarray}
are all positive, and the minimum value of $N({\bf p})$ is zero. But
in this case the metric $\eta$ on the Hilbert space of the oscillators
become indefinite, which implies it is not always $+1$ but its general
form is $\eta=(-1)^{-N({\bf p})}$. In the present case we can define a
vacuum satisfying $N({\bf p})|0\rangle=0$. For such field excitations
with momentum ${\bf p}$ the Hamiltonian of a single oscillator can be
written as
\begin{eqnarray}
H({\bf p})=\frac12\epsilon({\bf p})\left[a({\bf p})\bar{a}({\bf p})
+\bar{a}({\bf p})a({\bf p})\right]=-\epsilon({\bf p})\left[N({\bf p})+
\frac12\,\delta^3({\bf 0})\right]\,,
\label{pham}
\end{eqnarray}
The additive term $-\frac12\,\epsilon({\bf p})\,\delta^3({\bf 0}) $ is
the zero point energy which can be ignored in further calculations. In
the present case the Hamiltonian of the above field excitations are
completely equivalent to the negative energy Hamiltonian of the Pauli
oscillator in Ref.~\cite{Pauli}.  Lee and Wick in
Ref.~\cite{Lee:1969fy,Lee} did show that it is indeed possible to
produce a sensible unitary quantum field theory with such kind of
excitations. In this article we are more interested about the thermal
distribution of the Lee-Wick or the Pauli excitations. Some one
interested in the formal structure of such kind of theories can refer
to the original work in Ref.~\cite{Lee:1969fy,Lee}.
Due to the presence of on-shell excitations the thermal vacuum becomes
$|\Omega\rangle\equiv |n({\bf p_1}),n({\bf p_2}),\cdots\rangle$ where
$n({\bf p_1})$ is the number of excitations carrying momentum ${\bf
  p_1}$. The action of the number operator on such a vacuum is
\cite{Lee}
\begin{eqnarray}
N({\bf p})|\Omega\rangle=n({\bf p})|\Omega\rangle,
\label{thermal-vacuum}
\end{eqnarray}
where $n({\bf p})$ is the number of particles with momentum ${\bf p}$
present in the thermal vacuum. In this analysis we will consider
non-interacting real scalar field for which the chemical potential
$\mu=0$.  In the present case with a Hamiltonian of the form as given
in Eq.~(\ref{pham}) the single particle partition function will be,
\begin{eqnarray}
z_{\rm LW}^B={\rm Tr}e^{-\beta H}=\sum_{n({\bf p})=0}^{\infty}e^{\beta n({\bf p})
\epsilon(\bf p)},
\label{zlw}
\end{eqnarray}
where $\beta=\frac{1}{T}$. From the last equation it is seen if we are
finding the single particle partition function for the Lee-Wick
partner of a Standard model boson then the series representing $z_{\rm
  LW}^{B}$ does not converge for $\beta > 0$. Consequently
we regularize the last expression by cutting off the summation for a
finite value of $n({\bf p})$ as:
\begin{eqnarray}
z_{\rm LW}^B=\sum_{n({\bf p})=0}^{M-1}e^{\beta n({\bf p}) \epsilon({\bf p})}=\frac{1 - e^{\beta 
\epsilon({\bf p}) M}}{1-e^{\beta \varepsilon({\bf p})}}\,,
\label{z2}
\end{eqnarray}
where $M$ is a phenomenological cut-off which can be made indefinitely
big at the end of the calculation. 

Next we calculate the thermal distribution function of the field
excitations, which behave like Pauli oscillator excitations, from the
expression of the single cell partition function of the field
excitations as given in Eq.~(\ref{z2}).  In conventional statistical
mechanics we can find the single cell distribution function via
\begin{eqnarray}
f({\bf p})=\frac{1}{\beta}\left(\frac{\partial \ln z}
{\partial \mu}\right)_{V,\beta}\,,
\end{eqnarray}
where $\mu$ is an auxiliary chemical potential whose exact nature is
not important for our purpose. In presence of an auxiliary chemical potential
the single particle partition function can be written as
\begin{eqnarray}
z_{\rm LW}^B=\sum_{n({\bf p})=0}^{M-1}e^{\beta n({\bf p})\{\epsilon({\bf p}) - \mu\}} =
\frac{1 - e^{\beta\left\{\epsilon({\bf p}) - \mu\right\}M}}
{1-e^{\beta \left\{\epsilon({\bf p}) - \mu\right\}}}\,.
\end{eqnarray}
Applying conventional methods, the distribution function can
also be written as
\begin{eqnarray}
f_{\rm B}({\bf p})=\frac{1}{\beta}\left(\frac{\partial \ln z^B_{\rm LW}}
{\partial \mu}\right)_{V,\beta}\,,
\label{LW_stat}
\end{eqnarray}
which comes out to be,
\begin{eqnarray}
f_{\rm B}({\bf p})=
-\frac{e^{\beta\left\{\epsilon({\bf p})-\mu\right\}}}
{1-e^{\beta\left\{\epsilon({\bf p})-\mu\right\}}} + 
M\frac{e^{\beta\left\{\epsilon({\bf p})-\mu\right\}M}}
{1-e^{\beta\left\{\epsilon({\bf p})-\mu\right\}M}} \,.
\end{eqnarray}
Now setting the auxiliary chemical potential to be zero, $\mu=0$, we
get the distribution function of the fields whose excitation spectra
is given by the Pauli oscillators as
\begin{eqnarray}
f_{\rm B}({\bf p})= -\frac{1}{e^{-\beta \epsilon({\bf p})}-1}
+ \frac{M}{e^{-\beta \epsilon({\bf p}) M}-1}\,.
\label{lwpb}
\end{eqnarray}
\begin{figure}[h!]
\centering
\includegraphics[width=13cm,height=8cm]{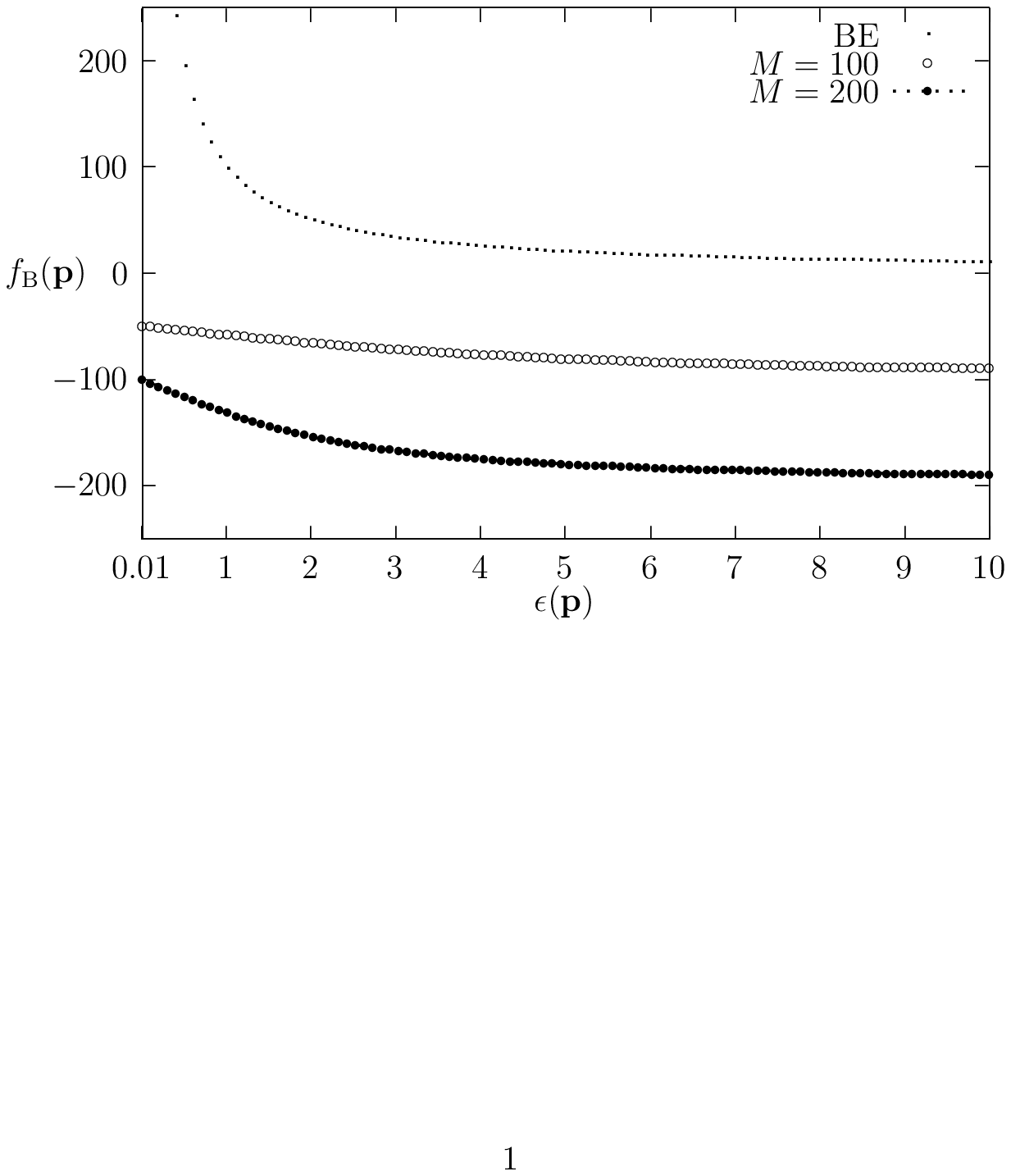}
\caption[]{The plot of the distribution function as given in
  Eq.~(\ref{lwpb}).  The topmost curve is for a normal Bose-Einstein
  distribution and the lower two curves correspond for $f_{\rm B}({\bf
    p})$ for the ultraviolet cutoff $M=100$ and $200$. The inverse
  temperature in all the cases is $.01{\rm GeV}^{-1}$ and the energy
  $\epsilon({\bf p})$ is in GeV.}
\label{lwb}
\end{figure}
This is the distribution function of the fields whose excitations are
described by creation and annihilation operators which satisfy
Eq.~(\ref{bquant}). These are not the fields which appear in the
Standard model of particle physics. The distribution function as
plotted in Fig.~\ref{lwb} shows that the average excitation per energy
level is negative definite. Obviously these systems describe a
physical theory which is non-trivial and the negative sign of the
distribution is only meaningful when compared with the positive
definite distribution of the normal Standard model bosons. In general
these kind of distributions will produce negative energy density and
pressure but once these energy density and pressure is added with the
positive energy density and pressure of the Standard model bosons we
get a net positive energy density and pressure. The important point to
notice about the distributions is that there is no pile up of quantas
near $\epsilon({\bf p})=0$ as is in the case of the Bose-Einstein
distribution. The reason being that the Pauli excitation spectra
distribution has two infinite spikes as the energy tends to zero and
they cancel each other near the origin.

If we assume the anticommutators of the creation and the annihilation
operators which define the fermionic excitations as
\begin{eqnarray}
a_s^2({\bf p})=\bar{a}_s^2({\bf p})=0\,,\,\,\,\,
\left\{a_s({\bf p})\,,\,\bar{a}_{s'}({\bf k})\right\}=-\delta_{s,s'}
\delta^3\left({\bf p}-{\bf k}\right)\,,
\label{fquant}
\end{eqnarray}
where $s$, $s'$ may be some internal quantum numbers, then we can
proceed in a similar way as done before and calculate the thermal
distribution of these excitations. If we stick to the definition of
the number operator as given in Eq.~(\ref{N}), then in this case the
possible eigenvalues of the number operator are simply 0 and 1. The
Hamiltonian of a single oscillator is
\begin{eqnarray}
H({\bf p})=\frac12\epsilon({\bf p})\left[a_s({\bf p})\bar{a}_s({\bf p})
-\bar{a}_s({\bf p})a_s({\bf p})\right]=-\epsilon({\bf p})\left[N({\bf p})-
\frac12\,\delta^3({\bf 0})\right]\,,
\label{phamf}
\end{eqnarray}
where $\frac12\,\epsilon({\bf p})\,\delta^3({\bf 0}) $ is the zero
point energy.  In the above equation we have assumed that the
Hamiltonian is independent of $s$. If we use this oscillator
Hamiltonian to calculate the single particle partition function,
neglecting the zero point energies, there is no problem regarding the
convergence of the series. The single particle partition function for
the anticommuting fields turns out to be
\begin{eqnarray}
z_{\rm LW}^F=\sum_{n({\bf p})=0}^{1}e^{\beta n({\bf p})\{\epsilon({\bf p}) - \mu\}}=
1+e^{\beta \left\{\epsilon({\bf p}) - \mu\right\}}\,,
\end{eqnarray}
where $\mu$ is an auxiliary chemical potential. Now applying the
formula in Eq.~(\ref{LW_stat}) and setting $\mu=0$ at the end we get
the distribution function for the Lee-Wick ghost partners or the Pauli
excitations of the Standard model fermions as:
\begin{eqnarray}
f_{\rm F}({\bf p})=-\frac{1}{e^{-\beta\epsilon({\bf p})}+1}\,.
\label{lwpf}
\end{eqnarray}
\begin{figure}[h!]
\centering
\includegraphics[width=13cm,height=8cm]{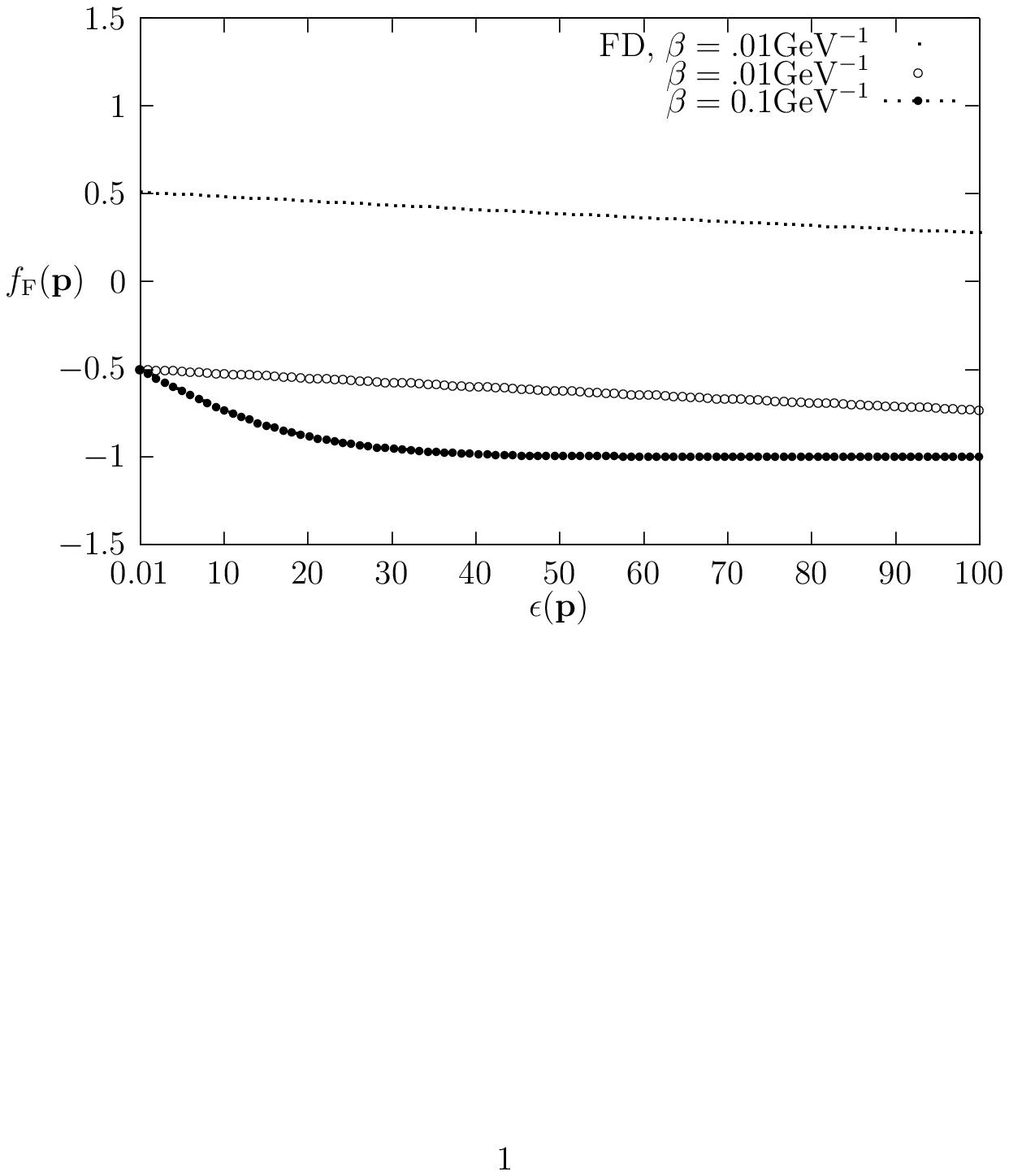}
\caption[]{The plot of the distribution function as given in
  Eq.~(\ref{lwpf}).  The topmost curve is for a normal Fermi-Dirac
  distribution at $\beta=.01{\rm GeV}^{-1}$ and the lower two curves
  correspond for $f_{\rm F}({\bf p})$ for $\beta$ values $.01{\rm
    GeV}^{-1}$ and $.1{\rm GeV}^{-1}$  and the energy
  $\epsilon{({\bf p})}$ is in GeV.}
\label{lwf}
\end{figure}
Unlike the previous case, in the present scenario the distribution
function has no dependence on the dimensionless regulator $M$. 

Here we point out that the fermionic degrees of freedom are not
exactly the same as presented in the paper by Lee and Wick. There they
assumed that the fermionic ghosts mix with each other and more over
they have complex masses. The description of the fermionic sector in
this article is more in line with the fermionic extension of Pauli's
work.  It is to be noted that although the Standard model particles
are proper bosons or fermions whose thermal distributions are given by
the Bose-Einstein or Fermi-Dirac distributions, in the present theory
the bosons (fields obeying commutation relations) or the fermions
(fields obeying anti-commutation relations) do not follow the Standard
Bose-Einstein or Fermi-Dirac statistics.

>From the nature of the distribution functions, for both the bosonic
and fermionic excitations, it is seen that the average number of
particles in an energy cell $\epsilon({\bf p})$ is negative. This is a
very counter intuitive result. But once we observe both the bosonic
and fermionic Hamiltonians we notice that the energy of any single
excitation of these fields are negative. The negative energy is not a
result of any particular choice of potential or interaction of the
fields but rather an artifact of the way these fields are
quantized. The vacuum defined is not stable and there exists much less
energetic states than the vacuum itself. These kinds of fields are
unstable. The maximum energy of the field configurations can be
zero. As the distribution functions are interpreted as average number
of excitations of the field quantas in any energy level $\epsilon({\bf
  p})$, which is positive, we immediately notice a contradiction. In
the present case the field configurations cannot afford to have
quantum excitations in positive energy states. Consequently the only
way there can be some positive energy out of these systems is by
removing some of the oscillator modes, or particles, from these
systems. As a result the negative distribution functions indicate that
the field configurations must be de-excited to get to positive energy
states i.e. the Lee-Wick partners decay into the Standard model particles.
\section{Energy density, pressure and entropy density from the distribution 
function.}
\label{eps}
\subsection{The bosonic case}
To calculate the relevant thermodynamic quantities for the bosonic
field from a statistical mechanical point of view we will employ
Eq.~(\ref{lwpb}). The energy density can be calculated using the
following known equation
\begin{eqnarray}
\rho = \frac{g}{\left(2\pi\right)^3}\int\epsilon({\bf p})f_{\rm B}
(\epsilon)d^3p
=\frac{g}{2\pi^2}\int_0^\infty \epsilon({\bf p})f_{\rm B}(\epsilon)
|{\bf p}|^2d{|\bf p}|\,,
\end{eqnarray}
where $g$ stands for any intrinsic degree of freedom of the particle.
For a relativistic excitation $\epsilon^2={\bf p}^2+m^2$ where $m$ is
the mass of the bosonic excitations. Changing the integration variable
from $|{\bf p}|$ to $\epsilon$ one gets
\begin{eqnarray}
\rho=-\frac{g}{2\pi^2}\int_0^\infty\left(\epsilon^3-\frac{m^2}{2}
\epsilon\right) \frac{d\epsilon}{e^{-\beta \epsilon}-1} + 
\frac{Mg}{2\pi^2}\int_0^\infty\left(\epsilon^3-\frac{m^2}{2}
\epsilon\right) \frac{d\varepsilon}{e^{-\beta\epsilon M}-1}\,.
\end{eqnarray}
In the above integral it is assumed that $|{\bf p}|\gg m$ and to have
a closed integral the lower limit of the integral is assumed to be
zero. In the extreme relativistic limit the system temperature $T \gg m$.
Both of the integrals can only be done when $\beta < 0$, and in that
case the result of the last integral is
\begin{eqnarray}
\rho=-\frac{g}{2\pi^2}\left(\frac{\pi^4T^4}{15}-
\frac{m^2\pi^2T^2}{12}\right) + \frac{g}{2\pi^2}
\left(\frac{\pi^4T^4}{15M^3}-
\frac{m^2\pi^2T^2}{12M}\right)\,.
\end{eqnarray}
Analytically continuing the above result for $\beta>0$ and taking $M
\to \infty$ we see that for normal temperatures the energy density for
extreme relativistic excitations of the bosonic fields is of the
following form
\begin{eqnarray}
\rho= -g\left(\frac{\pi^2T^4}{30}-\frac{m^2T^2}{24}\right)\,.
\label{endenb}
\end{eqnarray}
As expected, the energy density turns out to be negative for the
excitations in this case. The pressure density of the bosonic field
excitations can be found out from
\begin{eqnarray}
p &=&\frac{g}{\left(2\pi\right)^3}\int\frac{|{\bf p}|^2}{3\epsilon}
f_{\rm B}(\epsilon) d^3p=\frac{g}{2\pi^2}\int_0^\infty
\frac{|{\bf p}|^4}{3\epsilon}
f_{\rm B}(\varepsilon)d|{\bf p}|\,.
\end{eqnarray}
Following similar steps as in the case of the energy density, it is
seen that the pressure of extremely relavistic excitations of the
bosonic fields turns out to be
\begin{eqnarray}
p= -g\left(\frac{\pi^2T^4}{90}-\frac{m^2T^2}{24}\right)\,.
\label{pb}
\end{eqnarray}
The entropy density of the bosonic fields is simply given by
\begin{eqnarray}
s=\frac{\rho + p}{T} =-g\left(\frac{2\pi^2T^3}{45} - \frac{m^2T}{12}\right)\,.
\end{eqnarray}
These values of the energy density, pressure and entropy density
exactly match the corresponding values calculated for the Lee-Wick
partners, in \cite{Fornal}, in a different way. In \cite{Fornal} the
authors were trying to formulate thermodynamics for a
higher-derivative theory. The higher derivative theory was converted
into standard theory (theory up to a second derivative) with the
introduction of Lee-Wick partners. The authors in the previous work
did not quantize the system explicitly but were working with the form
of the propagators of the Lee-Wick partners.

If we assume that in the early universe for each bosonic degrees of
freedom in Standard model there exist a corresponding bosonic degree
of freedom whose creation and annihilation operators are quantized as
in Eq.~(\ref{bquant}) then the net energy density, pressure and
entropy density of the early universe turns out to be
\begin{eqnarray}
\rho_{B} = \rho_{\rm SM} + \rho = \frac{g m^2T^2}{24}\,,\,\,\,\,
p_{B} = p_{\rm SM} + p = \frac{g m^2T^2}{24}\,,\,\,\,\,
s_{B} = s_{\rm SM} + s=\frac{gm^2T}{12}\,,
\label{tboson}
\end{eqnarray}
 which are all positive as expected. Here the energy density, pressure
 density and entropy density for the Standard model bosonic particles
 are $\rho_{\rm SM}= g\frac{\pi^2T^4}{30}$, $p_{\rm
   SM}=g\frac{\pi^2T^4}{90}$ and $s_{\rm SM}=g\frac{2\pi^2T^3}{45}$
 respectively \cite{kolb}.
\subsection{The fermionic case}
In this subsection we apply Eq.~(\ref{lwpf}) to find the energy
density, pressure and entropy density of the fermionic excitations. In
this case the distribution function do not have any dependence on the
regulator $M$. For relativistic excitations the integrals which give
the energy density and pressure for the fermionic case are exactly
similar with the bosonic case except that now we have to use the
distribution for the fermions. The integrals can be easily done,
granted $\beta < 0$, but the results can be analytically continued for
positive temperatures. The results in this case are listed below. The
energy density, pressure and entropy density of the Lee-Wick partners
are as follows:
\begin{eqnarray}
\rho&=&-g\left(\frac{7\pi^2T^4}{240}-\frac{m^2T^2}{48}\right)\,,
\label{rhofi}\\
p &=&-g\left(\frac{7\pi^2T^4}{720}-\frac{m^2T^2}{48}\right)\,,
\label{pf}\\
s&=& -g\left(\frac{7\pi^2 T^3}{180} - \frac{m^2}{24}\right)\,.
\label{sfi}
\end{eqnarray}
The energy density and pressure quoted above was derived in
\cite{Fornal} in a different way for the special case of $g=2$. If we
assume that to each unusual fermionic degree of freedom there
corresponds one standard fermionic degree from the Standard model,
then the total fermionic contribution is
\begin{eqnarray}
\rho_F = \rho_{\rm SM} + \rho = \frac{g m^2T^2}{48}\,,\,\,\,
p_F = p_{\rm SM} + p = \frac{g m^2T^2}{48}\,,\,\,\,
s_F = s_{\rm SM} + s = \frac{g m^2T}{24}\,
\label{sf}
\end{eqnarray}
which are also all positive. Here the energy density, pressure density
and entropy density for the Standard Model fermionic particles are $\rho_{\rm
  SM}=g\frac{7\pi^2T^4}{240}$, $p_{\rm SM}=g\frac{7\pi^2T^4}{720}$ and
$s_{\rm SM}=g\frac{7\pi^2T^3}{180}$ respectively \cite{kolb}.

It is worth pointing out here that higher derivative theories of
fermions require two auxiliary Lee-Wick partners (one left-handed and
the other right-handed) to eliminate the higher derivative terms. In
that case the Lee-Wick degrees of freedom exceeds the one of its
Standard model partner yielding negative energy, pressure and entropy
density \cite{Wise}. This issue is yet to be resolved.
\section{Discussion and conclusion}
\label{conc}
In this article we have studied a system of bosonic and fermionic
fields, whose excitations modes are quantized with the negative sign
of the commutator or anticommutator. The bosonic part of the theory
was well studied before by Pauli \cite{Pauli} and the fermionic part
is studied previously by Lee and Wick \cite{Lee}. These unusual
quantization process naturally produces field configurations whose
total energy is negative. The negative energy of the field
configuration is not due to any particular form of the potential but
solely an outcome of the quantization process. The vacuum of the
theory is not the state with the lowest energy, it is rather the state
with the maximal energy making the field configuration unstable. The
bosonic and fermionic degrees of freedom do still follow commutation
and anticommutation relations and specifically the fermionic fields
still follow the Pauli exclusion principle. In this article the
emphasize had been on the calculation of energy density, pressure and
entropy density of the unusual filed configurations. To calculate the
above mentioned thermodynamic quantities one requires to have a
statistical mechanics of the field excitations. One encounters the
difficulty of a diverging sum when calculating the single particle
partition function of the bosonic fields. Keeping to conventional
ways, where the temperature of the system is positive definite, the
partition function can only be summed when one uses a ultraviolet
cutoff. The distribution function calculated from the partition
function turns out to be negative definite, which is a nontrivial
result. The negative nature of the distribution function implies that
there must be an average loss of particles in any positive energy
level. This fact can be understood by noticing that the system can
only have positive energy by loosing negative energy quanta and the
negative sign of the distribution function implies such a condition. 

It is shown in the article that the entropy density of the non-trivial
field configurations turn out to be negative. This is a serious issue
as by its very nature entropy is always positive. The problem with the
negative entropy of the Lee-Wick partners is avoided if we include the
entropy of the Standard model particles also. The total bosonic or
fermionic entropy turns out to be positive. We presume that the
negative entropy of the Pauli excitations and Lee-Wick partners show
that they cannot exist alone, the thermodynamic system is only
complete if the Standard model particles are also included.

The energy density, pressure calculated from the distribution
functions of the unusual fields discussed in this article are exactly
the same as calculated by Fornal, Grinstein and Wise in \cite{Fornal}.
The derivation of the new distribution functions and using them to
show the apparent connection between the Pauli oscillator
thermodynamics and the thermodynamics of the Lee-Wick partners are
the main motivations for this work.  Apparently the work of the
previous authors and the work in the present article has a common
conceptual similarity. The higher derivative theories and their lower
derivative Lee-Wick partner infested theories were motivated to kill
the divergences in a quantum field theory. The original work of Pauli
\cite{Pauli} also aimed to produce a quantum field theory devoid of
divergences. We strongly believe that the work of Pauli and Grinstein,
O'Connell and Wise have an underlying equivalence although the two
theories arises from different perspectives.  In short, the present
article acts as a bridge between higher derivative theories and normal
theories with deformed quantization conditions.  We hope that similar
to the Lee-Wick theory, there can be a theory of the early universe
where all the normal fields are accompanied by Pauli bosons or
fermions which are quantized with a negative commutator or
anticommutator bracket. This kind of a theory has the potential of
damping the infinite divergences which crop up in normal quantum field
theories. The unusual field configurations discussed in this article
may have been present in the very early universe and, like the
Lee-Wick ghosts, have decayed to other particles as time went on.
\vskip .2cm
\noindent 
{\bf Acknowledgement}: We want to thank  Subhendra Mohanty
for pointing out some subtle ideas involved in this article.


\begin{thebibliography}{999}
\bibitem{Lee:1969fy}
T.~D.~Lee and G.~C.~Wick,
Nucl.\ Phys.\  B {\bf 9}, 209 (1969).

\bibitem{Lee}
T.~D.~Lee and G.~C.~Wick,
Phys.\ Rev.\  D {\bf 2}, 1033 (1970).


\bibitem{Grinstein}
B.~Grinstein, D.~O'Connell and M.~B.~Wise,
Phys.\ Rev.\  D {\bf 77}, 025012 (2008)
[arXiv:0704.1845 [hep-ph]].

\bibitem{Carone:2008bs}
  C.~D.~Carone and R.~F.~Lebed,
  Phys.\ Lett.\  B {\bf 668}, 221 (2008)
  [arXiv:0806.4555 [hep-ph]].

\bibitem{Cai:2008qw}
Y.~F.~Cai, T.~t.~Qiu, R.~Brandenberger and X.~m.~Zhang,
Phys.\ Rev.\  D {\bf 80}, 023511 (2009)
[arXiv:0810.4677 [hep-th]].

\bibitem{Fornal}
B.~Fornal, B.~Grinstein and M.~B.~Wise,
Phys.\ Lett.\  B {\bf 674}, 330 (2009)
[arXiv:0902.1585 [hep-th]].

\bibitem{Dashen}
R.~Dashen, S.~K.~Ma and H.~J.~Bernstein,
Phys.\ Rev.\  {\bf 187}, 345 (1969).


\bibitem{Pauli}
W.~Pauli,
Rev.\ Mod.\ Phys.\  {\bf 15}, 175 (1943).


\bibitem{kolb} Kolb \& Turner, {\it The Early Universe} (Westview
  Press, 1994) Chapter: 3

\bibitem{Wise}
  M.~B.~Wise,
  Int.\ J.\ Mod.\ Phys.\  A {\bf 25}, 587 (2010)
  [arXiv:0908.3872 [hep-ph]].
 
\end{thebibliography}
\end{document}